\documentclass{myws-p8}
\usepackage{graphicx}
\usepackage{psfrag,here}

\def\al{&}
\def\be{\begin{equation}}
\def\ee{\end{equation}}
\def\bea{\begin{eqnarray}}
\def\eea{\end{eqnarray}}

\newcommand{\bdm}{\begin{displaymath}}
\newcommand{\edm}{\end{displaymath}}
\newcommand{\no}{\nonumber \\}

\newcommand{\fs}{\; .}
\newcommand{\co}{\; ,}
\newcommand{\eff}{{e\hspace{-0.1em}f\hspace{-0.18em}f}}
\newcommand{\QCD}{{\mbox{\tiny Q\hspace{-0.05em}CD}}}

\newcommand{\lvac}{\langle 0|\,}
\newcommand{\rvac}{\,|0\rangle}
\newcommand{\nc}{N_{\!c}}
\newcommand{\qbar}{\overline{\rule[0.42em]{0.4em}{0em}}\hspace{-0.45em}q}
\newcommand{\ubar}{\overline{\rule[0.42em]{0.4em}{0em}}\hspace{-0.5em}u}
\newcommand{\dbar}{\,\overline{\rule[0.65em]{0.4em}{0em}}\hspace{-0.6em}d}

\newcommand{\ind}{\scriptscriptstyle}
\newcommand{\R}{{\mbox{\tiny R}}}
\renewcommand{\L}{{\mbox{\tiny L}}}
\newcommand{\nf}{N_{\!f}}

\begin{document}

\title{Theoretical Chiral Dynamics}

\author{H.~Leutwyler}

\address{Institute for Theoretical Physics, University of Bern\\
Sidlerstr.~5, CH-3012 Bern, Switzerland\\E-mail: leutwyler@itp.unibe.ch}

\maketitle

\abstracts{The reasons why a considerable effort is made to resolve  
the low energy structure of QCD are discussed.
The effective field theory used for this purpose is illustrated with the 
recent progress made in the
predictions for $\pi\pi$ scattering and in understanding the low energy
properties of the theory in the large $\nc$ limit.}

\vspace*{-1.5em}
\begin{center}{ \normalsize\it Talk given at Chiral 2000, 
Jefferson Laboratory, \\Newport News, Virginia, July 2000.}
\end{center}

\section{Standard Model}
In the Standard Model, the dynamical variables are the gauge bosons
$\gamma, W,Z,G$, the Higgs fields $(\phi_1,\phi_2,\phi_3,\phi_4)$, the
quarks $q$ and the leptons $\ell$. Except for the mass term of the
Higgs field, the Lagrangian does not contain mass parameters -- the masses of
the various particles are of dynamical origin:
The ground state contains a condensate of neutral Higgs particles,
$\lvac \phi_1\rvac\neq 0$. Neither the
photon nor the gluons take notice -- for these, the vacuum is transparent, 
because $\phi_1$ is electrically neutral and does not carry colour. 
For the gauge
fields that mediate the weak interaction, however, this is not the case:
The vacuum is not transparent for $W$ and $Z$ waves of low frequency -- these
particles do interact with those forming the condensate, because $\phi_1$ is 
not neutral 
with respect to flavour. As a consequence, the frequency of the  
$W$ and $Z$ waves tends to a nonzero value at large wavelength: The
corresponding 
particles move at a speed that is smaller than the velocity of light -- both 
the $W$ and the $Z$ pick up a mass. 

The quarks and leptons also interact 
with the particles in the condensate and thus also pick up mass.  
It so happens that the interactions of $\nu,e,\mu,u,d,s$ with the Higgs fields
are weak, so that the masses $m_\nu,m_e,m_\mu,m_u,m_d,m_s$ are small. 
The remaining fermion masses, as well as $m_W$, $m_Z$ and $m_H$ 
are not small. We do not know why the observed mass pattern looks like this,
but we can analyze the consequences of this empirical fact. 

At energies that are small compared to $\{m_W,m_Z,m_H\}=O(100\,\mbox{GeV})$, 
the weak interaction freezes out, because these
energies do not suffice to bridge the mass gap and to excite the corresponding
degrees of freedom.  
As a consequence, the gauge group of the Standard Model, 
SU(3)$\times$SU(2)$\times$U(1), breaks down to
the subgroup SU(3)$\times$U(1) -- only the photons, the gluons, the quarks and
the charged leptons are active at low energies. Since the neutrini
neither carry colour nor charge, they decouple. 

\section{Effective theory for {\boldmath
    $E\ll 100\,\mbox{GeV}$\unboldmath}} 

The Lagrangian relevant in the low energy domain is the one of QCD $+$ QED,
which is characterized by the two
coupling constants $g$ and $e$.  
In contrast to the Standard Model,
the SU(3)$\times$U(1) Lagrangian does contain mass terms: the quark and lepton
mass matrices $m_q$, $m_\ell$. 
Moreover, Lorentz and gauge invariance
permit the occurrence of a term 
proportional to the operator
\bea\label{omega} \omega=
\frac{1}{16\,\pi^2}\,\mbox{tr}\hspace{-0.6em}
\rule[-0.5em]{0em}{0em}_c\hspace{0.5em} G_{\mu\nu}\,\tilde{G}^{\mu\nu}\fs
\eea
The corresponding coupling constant $\theta$ is referred to as 
the vacuum angle. 
The field basis may be chosen such that $m_q$ and $m_\ell$ are
diagonal and positive.
The fact that the electric dipole moment of the
neutron is very small implies that -- in this basis -- 
$\theta$ must be tiny. This is called the strong CP-problem: We do not
really understand why the 
neutron dipole moment is so small.

The two gauge fields involved in the effective low energy theory behave in a
qualitatively different manner: While the photons do not carry electric
charge, the gluons do carry colour. This difference is responsible for the fact
that the strong interaction becomes strong at low energies, while the
electromagnetic interaction becomes weak there, in fact remarkably weak:
The photons and leptons essentially decouple from the quarks and gluons.
The electromagnetic interaction can be accounted for by means of the
perturbation series in powers of $e$. For the QCD part of the theory, on the
other hand, perturbation theory is useful only at high energies. In the low
energy domain, the strong interaction is so strong that it 
confines the quarks and gluons.

The resulting effective low energy theory is mathematically more 
satisfactory than the 
Standard Model as such -- it does not involve scalar degrees of freedom and
has fewer free parameters. Remarkably, this simple theory 
must describe the
structure of cold matter to a very high degree of precision, once 
the parameters in the Lagrangian are known. It in particular 
explains the size of the atoms in terms of the scale
\bea a_{\hspace{-0.07em}\ind B}=\frac{4\,\pi}{e^2\,m_e}\co\nonumber\eea 
which only contains the two parameters $e$ and $m_e$ -- these are indeed known
to an incredible precision. Unfortunately, our ability to solve the QCD part
of the theory is rather limited -- in particular, we are
still far from being able to demonstrate on the basis of the QCD Lagrangian
that the strong interaction actually
confines colour. Likewise, our knowledge of the magnitude of the light quark
masses is still rather limited -- we need to know these more accurately 
in order to test ideas that might lead to an understanding of the
mass pattern, such as the relations with the lepton masses that emerge from 
attempts at unifying the electroweak and strong forces.

\section{Massless QCD -- a theoretical paradise}
In the following, I focus on the QCD part and switch the electromagnetic
interaction off.  As mentioned already, $m_u,m_d$ and $m_s$ happen to be
small. Let me first set these parameters equal to zero and, moreover, 
send the masses of the heavy quarks,
$m_c,m_b,m_t$ to infinity. In this limit, the theory becomes a theoreticians
paradise: The Lagrangian contains a single parameter, $g$. In fact, since 
the value of $g$ depends on the running scale used, the theory does not
contain any dimensionless parameter that would need to be adjusted to
observation. In principle, this theory fully specifies all dimensionless 
observables as pure numbers, while dimensionful quantities like masses or
cross sections can unambiguously
be predicted in terms of the scale $\Lambda_{\QCD}$ or the mass of the proton.
The resulting theory -- QCD with three massless flavours -- is among the
most beautiful quantum field theories we have. I find it breathtaking that, 
at low energies, nature reduces to this beauty, as soon as the dressing with
the electromagnetic interaction is removed and the Higgs condensate is 
replaced by one that does not hinder the light quarks, but is impenetrable for
$W$ and $Z$ waves as well as for heavy quarks.

The Lagrangian of the massless theory, which I denote by ${\cal L}_{\QCD}^0$,
has a high degree of symmetry, which originates in the fact that the
interaction among the quarks and gluons is flavour-independent and conserves
helicity: ${\cal L}_{\QCD}^0$ is invariant under independent flavour rotations
of the three right- and left-handed quark fields. These form the group
$G=\mbox{SU(3)}_\R\times\mbox{SU(3)}_\L$. The corresponding 16 currents
$V^\mu_i\qbar\gamma^\mu\frac{1}{2}\,\lambda_i q$ and 
$A^\mu_i=\qbar\gamma^\mu\gamma_5\frac{1}{2}\,\lambda_i q$ are conserved, so
that their charges commute with the Hamiltonian:
\bea [\,Q_i^{\mbox{\tiny V}},H_\QCD^0\,]=
[\,Q_i^{\mbox{\tiny A}},H_\QCD^0\,]=0\co
\hspace{2em}i=1,\,\ldots\,,8\fs\nonumber\eea
Vafa and Witten~\cite{Vafa Witten} have shown that the state of lowest energy
is necessarily invariant under the vector charges: 
$Q_i^{\mbox{\tiny V}}\rvac=0$. For the axial charges, however, there are the
two possibilities characterized in table 1.
\begin{table}
\begin{tabular}{|c|c|}
\hline
\rule[-0.7em]{0em}{1.9em}$Q_i^{\mbox{\tiny A}}\rvac=0$&
$Q_i^{\mbox{\tiny A}}\rvac\neq0$\\ \hline\rule{0em}{1.2em}
Wigner-Weyl realization of $G$&Nambu-Goldstone realization of $G$ \\
ground state is symmetric & ground state is asymmetric\\
\rule[-1em]{0em}{2em}
$\lvac\qbar_\R q_\L\rvac = 0$&$\lvac\qbar_\R q_\L\rvac \neq 0$
\vspace*{-0.5em}\\
ordinary symmetry & spontaneously broken symmetry\\
spectrum contains parity partners & spectrum contains Goldstone bosons\\
\rule[-0.7em]{0em}{0em}degenerate multiplets of $G$& 
degenerate multiplets of $\mbox{SU(3)}\in G$\\
\hline\end{tabular}
\caption{Alternative realizations of the symmetry group
  $G=\mbox{SU(3)}_\R\times\mbox{SU(3)}_\L$.} 

\vspace*{-1em}
\end{table}

The observed spectrum does not contain parity doublets. In the case
of the lightest meson, the $\pi(140)$, for instance, the lowest state with 
the same spin and flavour quantum numbers, but opposite parity is the 
$a_0(980)$. So, expe\-ri\-ment rules out the first possibility. 
In other words, 
for dynamical reasons that yet remain to be understood,
the state of lowest energy is an asym\-metric state. 
Since the axial charges
commute with the Hamiltonian, there must be eigenstates with the same energy
as the ground state:
\bea H^0_\QCD\, Q_i^{\mbox{\tiny A}}\rvac= Q_i^{\mbox{\tiny A}}\,H^0_\QCD\rvac
=0\fs\nonumber\eea
The spectrum must contain 8 states $Q_1^{\mbox{\tiny A}}\rvac,\ldots\,,
Q_8^{\mbox{\tiny A}}\rvac$  
with $E=\vec{P}=0$, describing massless particles, the Goldstone bosons of the
spontaneously broken symmetry. Moreover, these must carry spin 0, negative
parity and form an octet of SU(3). 
 
\section{Quark masses as symmetry breaking parameters}

Indeed, the 8 lightest hadrons,
$\pi^+,\pi^0,\pi^-,K^+,K^0,\bar{K}^0,K^-,\eta$,
do have these quantum numbers, but massless they
are not.
This has to do with the deplorable fact that we are not living in paradise: The
masses $m_u,m_d,m_s$ are different from zero and thus allow the left-handed
quarks to communicate with the right-handed ones. 
The Lagrangian is of the form
\bea {\cal L}_\QCD={\cal L}_\QCD^0-\qbar_\R m\, q_\L-
\qbar_\L m^\dagger q_\R\co\hspace{2em}
m=\left(\!\!\!\mbox{\begin{tabular}{ccc}\vspace*{-0.5em}$m_u\!\!\!$&
&\\\vspace*{-0.5em}&$\!\!\!m_d\!\!\!
$&\\&&$\!\!\!m_s$
\end{tabular}}\!\!\!\!\right)\fs\nonumber\eea
The quark masses may be viewed as symmetry breaking parameters: The 
QCD-Hamiltonian is only approximately symmetric under independent rotations of
the right- and left-handed quark fields, to the extent that these parameters
are small. Chiral symmetry is thus broken in two ways: 
\begin{itemize}\item spontaneously\hspace{3em} 
$\lvac\qbar_\R q_\L\rvac \neq 0$
\item explicitly\hspace{5.2em} $m_u,m_d,m_s\neq 0$\end{itemize}

The consequences of the fact that the explicit symmetry breaking is small may
be worked out by means of an effective field theory.~\cite{Weinberg
Physica,reviews,BCE} The various quantities of interest are expanded in powers 
of the momenta and quark masses. In the case of the pion mass, for instance,
the expansion starts with~\cite{GMOR}  
\bea\label{Mpi} M_{\pi^+}^2=(m_u+m_d)\,B_0+O(m^2)\co
\hspace{2em}B_0=\frac{1}{F_0^2}\,|\lvac 
\ubar u \rvac|\fs\eea
$F_0$ is the value of the pion decay constant in the chiral limit,
$m_u,m_d,m_s\rightarrow 0$.
The formula shows that the square of the pion mass is proportional to the
product of $m_u+m_d$ with the order parameter $\lvac 
\ubar u \rvac$. The two factors represent quantitative measures for explicit 
and spontaneous symmetry breaking, respectively. If the explicit symmetry
breaking is turned off, the pions do become massless, as they
should: The symmetry is then exact, so that 
the spectrum must contain a massless Goldstone boson octet, while all 
other levels form massive, degenerate multiplets of SU(3). Actually,
the excited
mesonic states are unstable, because the strong interaction allows them 
to decay into the Goldstone bosons, but the symmetry ensures that both the
mass and the lifetime
of the members of a given multiplet are the same.

\section{Illustration: {\boldmath $\pi\pi$\unboldmath} scattering lengths}
Above, I treated the masses of all three light quarks as expansion
para\-meters. For the
low energy analysis of the $\pi\pi$ scattering amplitude, however,
there is no need to
expand in powers of $m_s$. We can keep $m_s$ at its physical value and
only expand in powers of $m_u$ and $m_d$. In the limit $m_u,m_d\rightarrow 0$
at fixed $m_s$, QCD already acquires an exact symmetry: The Hamiltonian
becomes invariant under the group SU(2)$_\R\times$SU(2)$_\L$ of chiral
rotations in the space spanned by the two massless flavours. The ground state
spontaneously breaks that symmetry to the subgroup SU(2)$_{\mbox{\tiny V}}$ --
the good old isospin symmetry discovered in the thirties of the last
century.~\cite{Heisenberg} 
Only the pions then become massless, while the kaons and the $\eta$ remain
massive. In the following, I consider this framework and ignore isospin
breaking, setting 
$m_u=m_d=\hat{m}$. The partial wave decomposition then contains two
independent $S$-waves, corresponding to $s$-channel isospin $I=0$ and $I=2$
(Bose statistics does not permit an $S$-wave with $I=1$).
The corresponding scattering lengths are denoted by $a_0^0$ and $a_0^2$ -- the
lower index specifies the total angular momentum, while the upper one refers
to isospin.

As a general consequence of the symmetry, Goldstone bosons of zero momentum
cannot interact with one another: $a_0^0$ and $a_0^2$ vanish for
$m_u,m_d\rightarrow 0$. These quantities thus also measure the explicit
symmetry breaking generated by the quark masses, like $M_\pi^2$. In fact, 
Weinberg's low energy theorem~\cite{Weinberg 1966} states that, 
to leading order of the expansion in powers of $m_u$ and $m_d$,
the scattering lengths are proportional to $M_\pi^2$, the factor of
proportionality being fixed by the pion decay constant. The low energy theorem
may be written in the form\footnote{The
standard definition of the scattering length 
corresponds to $a_0/M_\pi$. It is not suitable in the present context, 
because it differs from the invariant 
scattering amplitude at threshold by a kinematic factor that diverges in the
chiral limit.}
\bea a_{0}^0=\frac{7 M_\pi^2}{32 \,\pi \, F_\pi^2}\,R_0\co
\hspace{1.3em}
a_{0}^2=-\frac{M_\pi^2}{16 \,\pi \,
  F_\pi^2}\,R_2\co\hspace{1.3em}R_I=1+O(\hat{m})\fs\eea
The two loop representation~\cite{BCEGS} explicitly specifies the
scattering 
lengths in terms of the effective coupling constants, up to and including
contributions of $O(\hat{m}^3)$, so that the correction factors 
$R_0,R_2$ can be
calculated to next-to-next-to leading order. Explicitly, the result
reads~\cite{Colangelo 1995,Colangelo 1997}
\bea\label{R0R2} R_0\al=\al 1+x\,\frac{9}{2}\,
\ln \frac{M_{0}\!\rule{0em}{0.8em}^2}{M\rule{0em}{0.7em}^2}\,+x^2\,
\frac{769}{84}\left(\ln
  \frac{\tilde{M}_{0}\!\rule{0em}{0.8em}^2}{M\rule{0em}{0.7em}^2} 
\right)^{\!\!2}+x^2\,k_0+O(x^3)\co\\
R_2\al=\al  1-x\,\frac{3}{2}
\ln \frac{M_{2}\!\rule{0em}{0.8em}^2}{M\rule{0em}{0.7em}^2}\,- x^2\,
\frac{17}{12}\left(\ln \frac{\tilde{M}_{2}\!\rule{0em}{0.8em}^2}
{M\rule{0em}{0.7em}^2}
\right)^{\!\!2}+x^2\,k_2+O(x^3)\co\nonumber\eea
where $x\propto\hat{m}$ stands for the dimensionless quantity
\bea x =\left(\!\frac{M}{4\hspace{0.05em} \pi
    F}\!\right)^{\!\!2}\co \eea
which measures the pion mass in units of the pion
decay constant. More precisely, $M^2= 2\, \hat{m}\, B$ is the first 
term in the expansion of $M_\pi^2$ 
and $F$ is the value of the pion
decay constant for $m_u=m_d=0$. Note that $B$ and $F$ differ from the
quantities $B_0$ and $F_0$ in equation 
(\ref{Mpi}) by contributions of $O(m_s)$.

\section{Infrared singularities}
The expansion in powers of the quark masses may be obtained by splitting 
the QCD Hamiltonian into two parts,
\bdm H_\QCD=H_0+H_1\co\hspace{1.3em}
H_1=\!\int\!d^3\!x\{m_u\,\ubar u+m_d \,\dbar d\}\co 
\edm
and treating $H_1$ as a perturbation. 
The formula (\ref{R0R2})
shows that the resulting expansion of the scattering lengths
is not an ordinary Taylor series, but contains terms involving the logarithm
thereof. This indicates that the straightforward
perturbation series in powers of $H_1$ runs into infrared singularities, which
occur because the spectrum of $H_0$ contains massless particles,
pions. The infrared divergences are of logarithmic type. The coefficients
of the logarithms are determined
by the structure of the symmetry group and the transformation properties of
$H_1$. 

In the case of $R_0$, the coefficients of the
logarithms are positive and unusually large, while in the case of $R_2$ they
are negative and of normal size. The qualitative difference can be understood
on the basis of unitarity, which requires the partial waves to contain
a branch cut for $s\geq 4 M_\pi^2$, generated by the final state interaction. 
In the $I=0$ channel, 
this interaction is attractive and very strong, while for $I=2$, 
it is repulsive and weak, because that channel is exotic.
As the scattering lengths represent the values of the partial waves at the
branch point, we are in effect considering the expansion of the scattering
amplitude at a point where it is singular. The fact that the
coefficients of the logarithms occurring in the expansion of $R_0$ are
extraordinarily large implies that, in this case, the series converges 
unusually slowly. 

The logarithmic scales are not determined by the symmetry.
In the language of the effective theory, these scales are related
to the values of the effective coupling constants. Indeed, not only the
scales $M_0,M_2$ that specify the corrections of order $\hat{m}$, 
but also those of the terms proportional to the square of a
logarithm, are determined by the effective
coupling constants that occur at first
nonleading order in the derivative expansion of the effective Lagrangian,
${\cal L}_\eff={\cal L}^{(2)}+{\cal L}^{(4)}+{\cal L}^{(6)}+\ldots\,$
It is convenient to use the corresponding running couplings at the scale
$\mu=M$, which are denoted by~\cite{GL 1984} 
$\bar{\ell}_1,\ldots,\bar{\ell}_4$ and
also depend logarithmically on $M$:
\bea\label{Lambda} \bar{\ell}_n\al=\al \ln 
\frac{\Lambda_n\!\rule{0em}{0.8em}^2}{M\rule{0em}{0.7em}^2}\fs\eea
In terms of these quantities, the scales occurring in the expansion of the
scattering lengths are given by
\bea
\ln
\frac{M_{0}\!\rule{0em}{0.8em}^2}{M\rule{0em}{0.7em}^2}\al=\al\frac{1}{189}\,
(40\,\bar{\ell}_1+80\,\bar{\ell}_2-15\,\bar{\ell}_3+
84\,\bar{\ell}_4)+\frac{5}{9}\co\no
\ln
\frac{M_{2}\!\rule{0em}{0.8em}^2}{M\rule{0em}{0.7em}^2}\al=\al\frac{1}{9}\,
(8\,\bar{\ell}_1+16\,\bar{\ell}_2-3\,\bar{\ell}_3-
12\,\bar{\ell}_4)+\frac{1}{3}\co\\
\ln
\frac{\tilde{M}_{0}\!\rule{0em}{0.8em}^2}{M\rule{0em}{0.7em}^2}
\al=\al\frac{1}{769}\,
(132\,\bar{\ell}_1+568\,\bar{\ell}_2-267\,\bar{\ell}_3+
336\,\bar{\ell}_4)+\frac{1361}{1538}\co\no
\ln
\frac{\tilde{M}_{2}\!\rule{0em}{0.8em}^2}{M\rule{0em}{0.7em}^2}
\al=\al\frac{1}{17}\,
(-12\,\bar{\ell}_1+8\,\bar{\ell}_2-3\,\bar{\ell}_3+
24\,\bar{\ell}_4)-\frac{125}{34}\fs\nonumber
\eea 
The mass independent
terms $k_0$ and $k_2$ account for the remaining contributions of 
$O(\hat{m}^2)$, in
particular also for those from the effective couplings of 
${\cal  L}^{(6)}$.

\section{Numerical discussion}
Rough estimates for the coupling constants
$\bar{\ell}_1,\ldots\,,\bar{\ell}_4$ were given long 
ago.~\cite{GL 1984} In the meantime, the values of $\bar{\ell}_1$ and
$\bar{\ell}_2$ have been determined more 
accurately.~\cite{reviews,GKMS,Borges,Wanders} 
In order to analyze the corrections of order $\hat{m}^2$ in $R_0$ and $R_2$, 
we need phenomenological determinations of comparable accuracy --
one loop 
analyses are not suitable for the present purpose. One source of information,
where the analysis has now been done to two loops, is $K_{e_4}$
decay.~\cite{Bijnens Jefferson} The Roy equations for
$\pi\pi$ scattering~\cite{ACGL} allow an entirely independent
determination,~\cite{Colangelo Jefferson} 
which, moreover, only relies on SU(2)$_\R\times$SU(2)$_\L$.
The constant $\bar{\ell}_4$  
is related to the scalar charge radius, for which the calculation has also
been done to two loops.~\cite{BCT} The information
about $\bar{\ell}_3$, on the other hand, is still very meagre -- the value of
this constant makes 
the difference between the standard picture and ``Generalized Chiral
Perturbation Theory'' (see below). In the following, I invoke the very
crude estimate~\cite{GL 1984} 
$\ln \Lambda_3\rule{0em}{0.8em}^{\!\!2}/M_\pi\rule{0em}{0.8em}^{\!\!2}
=2.9\pm 2.4$, which amounts to $0.2\, \mbox{GeV}<\Lambda_3<2\,\mbox{GeV}$.
Concerning the values of $k_0$ and $k_4$, I rely on the
estimates for the coupling constants of ${\cal L}^{(6)}$
in the literature.~\cite{BCEGS,BCT} 
\begin{figure}[thb] 
\psfrag{I0}{\hspace{0.2em}\raisebox{1.3em}{$R_0$}}
\psfrag{I2}{\raisebox{0.7em}{$R_2$}}
\psfrag{a}{}
\psfrag{m}{$_{\rule{0.5em}{0em}\textstyle M_{\vspace{-0.08em}\pi}
\hspace{0.2em}\mbox{\small (MeV)}}$}
\psfrag{50}{$_{50}$}
\psfrag{100}{$_{100}$}
\psfrag{150}{$_{150}$}
\psfrag{200}{$_{200}$}
\psfrag{250}{$_{250}$}
\psfrag{300}{$_{300}\vspace{2em}$}
\psfrag{1}{\hspace{-0.3em}$1$}
\psfrag{2}{\hspace{-0.3em}$2$} 
\psfrag{3}{\hspace{-0.3em}$3$}

\vspace*{-1em}
\hspace*{6em}\mbox{\epsfysize=5cm \epsfbox{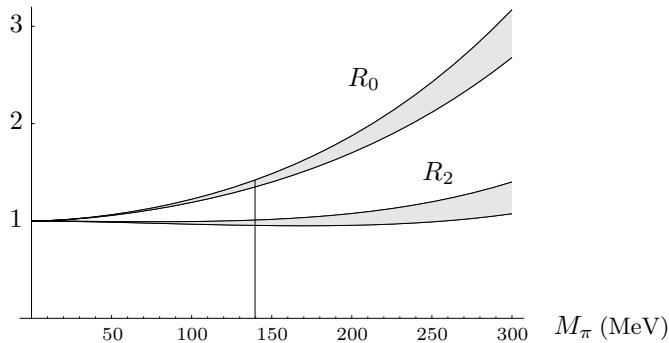} }

\caption{$\pi\pi$ $S$-wave scattering lengths as a function of the pion mass.} 

\end{figure}   

Fig.1 indicates the behaviour of the correction factors as functions of the
quark mass $\hat{m}$, which is varied from 0 (where the symmetry is exact) 
to about 20 MeV. The variable shown on the horizontal axis is the
corresponding pion mass. 
The figure shows that $R_0$ very rapidly grows
with the strength of the symmetry breaking. In fact, the 
chiral perturbation theory formulae underlying the figure are meaningful 
only in the range where the corrections are small 
(the shaded regions exclusively account for the 
uncertainties in the values of the coupling constants). If the quark masses 
are taken at their physical values, dropping the terms of 
$O(x^3)$ and using the central values for the coupling constants leads to
$R_0=1.36$. 

It is advantageous to replace the expansion of the
scattering amplitude at threshold by one in the unphysical region,  
where the series converges much more rapidly.~\cite{CGL} 
As discussed in detail by 
G.~Colangelo at this meeting,~\cite{Colangelo Jefferson}
that method yields a remarkably precise 
prediction for the scattering lengths: $a_0^0=0.220\pm0.005$,
$a_0^2=-0.0444\pm 0.0010$. In the language used above, these numbers
correspond to $R_0=1.38\pm 0.03$, $R_2=0.98 \pm0.02$. 
The lattice result,~\cite{lattice I2} 
$a_0^2=-0.0374\pm 0.0049$ or $R_2=0.82\pm0.11$, 
is on the low side, but not
inconsistent with the prediction. The current situation concerning the
scattering lengths is depicted in fig.2, which is taken from the reference
quoted above.~\cite{CGL}

\begin{figure}[thb] 
\leavevmode \begin{center}
\vspace*{-0.5em}\hspace*{-1em}
\mbox{\rotatebox{-90}{\epsfysize=8.1cm \epsfbox{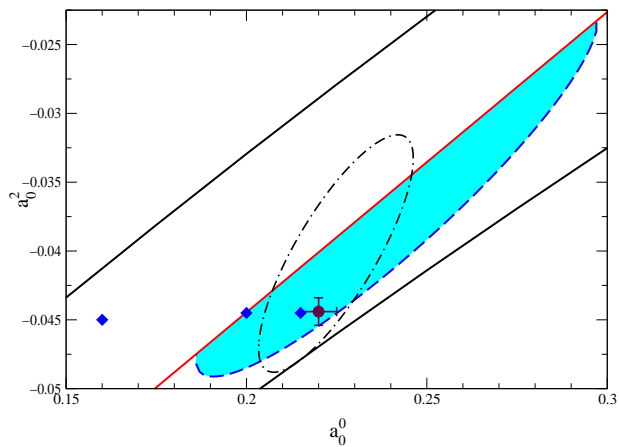} }}
\end{center}
\vspace*{-0.5em}
\caption{Scattering lengths: theory versus 
experiment.  
The shaded region represents the intersection of the 
domains allowed by the old data and by the Olsson sum rule.
The ellipse indicates the impact of the new, preliminary $K_{e_4}$
data. The three diamonds illustrate the
convergence of the chiral perturbation series at threshold 
(the one at the left corresponds to Weinberg's leading order formulae)
and the cross 
shows the result mentioned in the text.}
\end{figure}

\section{Is the quark condensate the leading order parameter ?}
The preceding discussion of the scattering lengths relies on the standard
hypo\-thesis, according to which the quark condensate represents the leading
order parameter of the spontaneously broken symmetry. This framework is
natural, because -- among the various operators that give rise to order
parameters -- $\qbar_\R q_\L$ is the one of lowest dimension. 
As emphasized by Stern and collaborators,~\cite{generalized} experimental
evidence for this to be the case is not available -- yet. 
The picture advocated by these authors may be motivated by an analogy with
spontaneous magnetization.
In that context, spontaneous symmetry breakdown
occurs in two quite different modes:
ferromagnets and antiferromagnets. For the former, the magnetization
develops a nonzero expectation value,
while for the latter, this does not
happen. In either case, the symmetry is spontaneosly broken. The example
illustrates that operators which the symmetry allows to pick up an 
expectation value may, but need not do so.

The issue concerns the validity of the Gell-Mann-Oakes-Renner relation,
equation 
(\ref{Mpi}). At first non-leading order and in the isospin limit 
$m_u=m_d=\hat{m}$, the correction is determined by the
coupling constant $\bar{\ell}_3$:
\bea M_\pi^2=2\,\hat{m}\,B-\frac{\hat{m}^2B^2}{8\pi^2
  F^2}\,\bar{\ell}_3+O(\hat{m}^3)\fs\eea
In this context, the standard hypothesis amounts to the assumption 
that the first term dominates over the second, so that $M_\pi^2$ is
approximately linear in $\hat{m}$. Since the structure of the ground state of
QCD is not understood, 
it is conceivable that the quark condensate is small, so that the
Gell-Mann-Oakes-Renner-relation fails, the ``correction'' of order 
$\hat{m}^2$ being comparable to or even larger than the first term. In the
language of the effective theory, this would imply that the estimate used
above for the coupling constant $\bar{\ell}_3$ is entirely wrong.
In the generalized framework, this constant is treated as a free parameter, so
that there is no prediction for the scattering lengths.

A sufficiently accurate measurement of the $\pi\pi$ scattering lengths would
decide the issue, because it amounts to a determination of the coupling
constant $\bar{\ell}_3$. An outcome like $a_0^0=0.26$, for example,
would be totally incompatible with the standard framework --
it would imply a value like $\bar{\ell}_3\simeq -20$. 
I am confident that the forthcoming results from Brookhaven,
CERN and Frascati will
provide a very significant test.

\section{QCD at large {\boldmath $N_c\!\!$ \unboldmath}}
As pointed out by 't Hooft,~\cite{tHooft} 
it is very instructive to vary the number of quark
colours, in particular to let it become large. The magnitude of the running
coupling $g$ of QCD is given by the familiar leading logarithmic formula
\bea \frac{g^2}{(4\pi)^2}=\frac{1}{\beta_0 \ln(
  {\mu^2}/{\Lambda_{\QCD}^2})}\co\hspace{2em}
\beta_0=\mbox{$\frac{1}{3}$}\,(11 \nc-2 \nf)\co\eea
where $\mu$ and $\Lambda_\QCD$ denote the running and intrinsic scales,
respectively. The formula shows that the coupling constant tends to zero in
proportion to $1/\sqrt{\nc}$ if $\nc$ is sent to infinity at a fixed value of
the ratio $\mu/\Lambda_\QCD$.  
This implies that graphs with the smallest possible number of quark loops then
dominate, so that the Okubo-Iizuka-Zweig rule becomes exact and the
constituent quark picture applies. In the large $\nc$ limit, the
scattering amplitudes are of $O(1/\nc)$ and the various
resonances become stable -- the width also shrinks in proportion to $1/\nc$.
In the following, I discuss a few consequences for the low energy properties
of the theory, drawing from work done in 
collaboration with 
R.~Kaiser.~\cite{Kaiser Leutwyler 2}  We are by no means the first to study the
subject -- a review may be found in that reference. 

The Ward identity obeyed by the singlet axial current contains an anomaly,
proportional to the operator $\omega$ defined in equation (\ref{omega}).
As is well-known, the anomaly term is suppressed when $\nc$ becomes large,
because it arises from graphs containing an extra quark loop:
In the large $\nc$ limit, 
the ninth axial current is conserved, so that
the theory acquires an additional symmetry, whose spontaneous breakdown gives
rise to a ninth Goldstone boson --
if the quark masses are turned off, the mass of the
$\eta'$ disappears when the number of colours is sent to infinity.

\section{Vacuum angle}
The correlation functions of $\omega$ are conveniently collected in the
effective action $S_\eff\{\theta\}$ that results if the QCD--Lagrangian is
perturbed with this operator,
\bdm {\cal L}_\QCD\rightarrow {\cal L}_\QCD-\theta\,\omega\fs\edm
In the present context, the $\theta$--term plays
a role of technical nature: $\theta=\theta(x)$ is treated as an external field
-- the value of physical interest is  $\theta=0$.

The expansion of the
effective action in powers of $1/\nc$ starts with a term of order $\nc^2$ that
arises from graphs without any quark lines. The large $\nc$ counting rules of
perturbation theory  
imply that the correlation function
$\lvac T\omega(x_1)\cdots \omega(x_n)\rvac$ is of $O(\nc^{2-n})$.
Accordingly, the leading term of the expansion depends on $\theta$
only through the ratio $\theta/\nc$:
\bea\label{S large N} S_\eff\{\theta\}=\nc^2\, S_0\{\vartheta\}+\nc\,
S_1\{\vartheta\}+\ldots\co\hspace{2em}\vartheta=\frac{\theta}{\nc}\fs
\eea 
The formula states that the dependence on the vacuum angle is suppressed.  
It is not
difficult to understand why that is so: The contribution from the
$\theta$--term is to be compared with the one from ${\cal L}^0_\QCD$. In the
relevant combination,
\bdm -\frac{1}{2g^2}\,\mbox{tr}\hspace{-0.6em}
\rule[-0.5em]{0em}{0em}_c\hspace{0.5em} \{G_{\mu\nu}\,G^{\mu\nu}+
\frac{g^2\theta}{8\,\pi^2}\, G_{\mu\nu}\,\tilde{G}^{\mu\nu}\}\co\edm
the weight of the $\theta$--term is smaller than the one that governs the
dynamics of the gluon field by $g^2\theta\propto \theta/\nc$.

It is important to notice that the large $\nc$ counting rules only hold for
generic momenta. The two-point-function, for instance,
\bdm i\! \int\!\! dx\, e^{i\,p\cdot x}\lvac T\omega(x)\omega(0)\rvac=
\frac{|\lvac \omega|\eta'\rangle|^2}{M_{\eta'}^2-p^2}+\ldots\edm
picks up a pole term from $\eta'$ exchange, which arises from graphs containing
at least one quark loop. If the quark masses are turned off, both $M_{\eta'}$ 
and $\lvac\omega|\eta'\rangle$ are quantities of $O(1/\sqrt{\nc})$,
so that the value of the pole
term at $p=0$ represents a contribution of $O(1)$, despite the  
counting rules, which state that, in $\lvac T\omega(x)\omega(0)\rvac$,
contributions from quark loops start showing up only at $O(1/\nc)$. 
If the vacuum angle is taken constant, we are in effect summing up 
the correlation functions of $\omega$ at zero momentum -- 
relations like (\ref{S large N}) are
not in general valid in that case. Also, there are paradoxical aspects 
in connection with periodicity in $\theta$ -- for a detailed discussion, I
refer to a paper written together with A.~Smilga.~\cite{Leutwyler Smilga}

\section{Effective theory at large {\boldmath $N_c$\unboldmath} and
  KM--transformation}
The standard form of the effective theory only accounts for the singularities
generated by the exchange of the particles contained in the pseudoscalar
octet. The contributions from all other states are described only
summarily, through their contributions to the effective coupling constants.
If the number of colours is allowed to become large, the effective theory must
be extended, introducing an additional field to describe the low energy
singularities due to $\eta'$ exchange. The effective Lagrangian then
contains a new low energy scale: $M_{\eta'}\propto 1/\sqrt{\nc}$. The low
energy structure of QCD can be analyzed within this extended framework, 
by means of a simultaneous expansion in powers of momenta, quark masses and
$1/\nc$. This machinery is particularly suited for analyzing the dependence
on the vacuum angle, because it explicitly accounts for the low energy
singularities that upset the large $\nc$ counting rules at exceptional momenta.
Indeed, the suppression of the $\theta$-dependence manifests itself in a
remarkable manner: To any finite order of
the $1/\nc$ expansion, the effective Lagrangian is a polynomial in the vacuum
angle, with coefficients that are suppressed by powers of $1/\nc$. 

As an illustration,\footnote{See also 
the contribution by R.~Kaiser in these proceedings.~\cite{Kaiser
  Jefferson}}
I briefly discuss the
ambiguity pointed out by Kaplan and Manohar.~\cite{Kaplan Manohar}  
The matrix
\bdm m' = m + \lambda\, e^{-i\theta} (m^+)^{-1} \det m^\dagger\co\edm
transforms in the same manner as the quark mass matrix $m$, under the full
group U(3)$_\R\times$U(3)$_\L$ of chiral rotations. Symmetry does 
therefore not distinguish $m'$ from $m$. Since the effective
theory exclusively exploits the symmetry properties of QCD, the above
transformation of the quark mass matrix does not change the form of the
effective Lagrangian relevant for $\nc=3$: The transformation may be 
absorbed in a suitable
change of the effective coupling constants, for any value of the vacuum angle,
in particular also for the case of physical interest, $\theta=0$.
This implies,
however, that the expressions for the masses of the pseudoscalars, 
for the scattering amplitudes or for the matrix elements of the vector and
axial
currents, which follow from this Lagrangian, are invariant under the operation
$m\!\rightarrow
\!m'$. Conversely, the experimental information on these observables
does not distinguish $m$ from $m'$. 

Since the KM-tranformation mixes the quark flavours,
it is evident that the parameter 
$\lambda$ violates the Okubo-Iizuka-Zweig rule and is therefore
suppressed in the large $\nc$ limit. Actually, 
a much stronger result can be established:~\cite{Kaiser Leutwyler 2} 
The transformation $m'\rightarrow m$ preserves the large 
$\nc$ properties of the theory only if $\lambda$ vanishes to all orders 
in $1/\nc$. This is a consequence of the conservation law
obeyed by the singlet axial current. The fact that the anomaly in this
conservation law disappears in the large $\nc$ limit implies that the
dependence on $\theta$ is suppressed. The KM--transformation is in conflict 
with this property, because the factor $e^{-i\theta}$ introduces a
dependence on $\theta$ that does not disappear when the number of colours
tends to infinity. For precisely the same reason,
an extra dynamical variable -- a field that describes the low energy
singluarities generated by the $\eta'$ -- is needed to cover the large $\nc$
limit: The constraint $\det U=e^{-i\theta}$, which is imposed on 
the meson field in the effective theory relevant for $\nc =3$, cannot be 
maintained in the
large $\nc$ limit.
 
In the remainder of the talk, I discussed a few aspects of chiral dynamics in
the context of the $\pi N$ interaction. That material is covered
in the contribution by T.~Becher.~\cite{Becher Jefferson} There are many
further developments in chiral dynamics, however,
which I did not have the time to discuss. For a more comprehensive picture,
I refer to the review articles listed in the bibliography.~\cite{reviews}

\section*{Acknowledgments}
It is a pleasure to thank the organizers of this workshop for a very pleasant
and stimulating meeting and the Swiss National Science Foundation for support.

\end{document}